\documentclass[lettersize,journal]{IEEEtran}
\usepackage{cite}
\usepackage{amsmath,amssymb,amsfonts}
\usepackage{algorithmic}
\usepackage{graphicx}
\usepackage{textcomp}
\usepackage{booktabs}
\usepackage{array}
\usepackage{makecell}
\usepackage{multirow}
\usepackage{graphicx}
\usepackage{color,soul}
\usepackage{tabularx}
\usepackage{lipsum}
\usepackage{multicol}
\usepackage{times}
\usepackage{fancyhdr,graphicx,amsmath,amssymb}
\begin{document}


\title{Blockchain Interoperability in UAV Networks: State-of-the-art and Open Issues}
\author{Ruba Alkadi, Noura Alnuaimi, Abdulhadi Shoufan, Chan Yeun
\thanks{ Center of Cyber-Physical Systems, Khalifa University, Abu Dhabi, UAE (emails: {ruba.alkadi, noura.alnuaimi, abdulhadi.shoufan, chan.yeun  }@ku.ac.ae)}}

\markboth{This work is submitted to an IEEE journal. }%
{Shell \MakeLowercase{\textit{et al.}}: A Sample Article Using IEEEtran.cls for IEEE Journals}


\maketitle

\begin{abstract}
The breakthrough of blockchain technology has facilitated the emergence and deployment of a wide range of Unmanned Aerial Vehicles (UAV) networks-based applications. Yet, the full utilization of these applications is still limited due to the fact that each application is operating on an isolated blockchain. Thus, it is inevitable to orchestrate these blockchain fragments by introducing a cross-blockchain platform that governs the inter-communication and transfer of assets in the UAV networks context. In this paper, we provide an up-to-date survey of blockchain-based UAV networks applications. We also survey the literature on the state-of-the-art cross blockchain frameworks to highlight the latest advances in the field. Based on the outcomes of our survey, we introduce a spectrum of scenarios related to UAV networks that may leverage the potentials of the currently available cross-blockchain solutions. Finally, we identify open issues and potential challenges associated with the application of a cross-blockchain scheme for UAV networks that will hopefully guide future research directions.

\end{abstract}


\section{Introduction}
\label{intro}
Recently, unmanned aerial vehicles (UAVs) have emerged as a game changing tech across many commercial industries. This fact is emphasized by the exponential increase in the UAV global market which is predicted to reach \$22.55 billion by the end of 2026 \cite{furtune}. The full deployment of UAVs' activities is, however, associated with safety, security, and reliability issues. These issues are fueling a surge of research activities to provide an optimum way to manage the airspace traffic flawlessly. Particularly, the Internet of Drones (IoD) concept has been resonating recently \cite{gharibi2016internet}. It basically fosters the idea of borrowing concepts from currently deployed networks (i.e. cellular networks, Air Traffic Management (ATM), and the Internet). Nonetheless, adopting concepts from these networks to the unmanned air traffic is not straightforward due to the heterogeneous nature of the latter. The UAVs' ability to move in three dimensions at high speeds makes the problem even harder. 

The blockchain technology, as a form of the Distributed Ledger Technology (DLT),  has proved effective in a multitude of security applications. Fundamentally, it is an immutable temper-proof distributed ledger that offers vital features such as treacability, transparency, and auditability. Together with its cryptographic algorithms, DLT serves as a secure repository of data and events. The blockchain works by first initiating a transaction where a set of nodes can verify or reject this transaction. The data in the blockchain is stored in blocks with specific sizes. Once a block reaches its maximum storage, it is linked to the block before it which will create a chain, hence the name blockchain. Due to its decentralized form, the blockchain offers traceability of the transactions. Blockchains can either be built permissioned or permissionless \cite{khan2019extended}. A permissioned blockchain requires the authentication of the nodes before processing any transactions. Nodes require authorization to be able to read and write data. A permissionless blockchain is, on the other hand, considered as a public blockchain where no authorization is required to read and write data. A blockchain can also be private or public, where a private blockchain can belong to a single company and all the nodes are controlled by that company while a public blockchain allows the public to join the network. 

The advantages of DLT were quickly recognized and thus applied in many fields to tackle various emerging problems. In this regard, it has been articulated that employing blockchain technology in the UAV networks context will mitigate security and safety risks and improve reliability. Alladi \textit{et al}. \cite{alladi2020applications} and Mehta \textit{et al}. \cite{mehta2020blockchain} have reviewed the literature on the deployment of blockchain for serving several UAV-based applications. This resulted in a large number of fragmented blockchains. Recently, it has been realized that these isolated chains need to communicate and inter-operate to exploit their full potential. The notion of cross-blockchain was proposed to address the portability and scalability of the blockchain technology. Portability refers to the ability of transferring assets and data between blockchains in a trustless way, while scalability refers to the ability to offload data to other blockchains. Several approaches to achieve interoperability between blockchains have fairly matured and successfully deployed for cryptocurrency exchanges. Besides, a multitude of applications have benefited from its potentials including healthcare data sharing \cite{med}, cyber-security \cite{neisse2020interledger}, video games \cite{besanccon2019towards}, and many others. Although the unmanned air traffic management industry could remarkably profit from the advances in cross-blockchain technology, none of the proposed blockchain-based UAV networks highlighted this potential. To the best of our knowledge, this work is the first to explore potential opportunities and use cases of cross-blockchain technology in the IoD context. Particularly, we are motivated by the enormous applications of the blockchain technology in the UAV networks and we believe that addressing the problem of interoperability of parallel blockchains is crucial to enabling efficient deployment of a decentralized immutable IoD environment. 

Mainly, our work aims at orchestrating current blockchain-based UAV networks by exploiting state-of-the-art cross-blockchain models. In brief, our contribution is distinguished by the following aspects:
\begin{itemize}
    \item Provide an up-to-date review of recent blockchain-based applications in an IoD environment. 
    \item Introduce possible scenarios where cross-blockchain is applicable in an IoD context.
    \item Propose a new cross-blockchain framework for UAV networks. 
\end{itemize}

 We, first, survey the literature on the current landscape of cross blockchain interoperability (Section \ref{Xchain}) and the latest blockchain-based UAV networks (section \ref{sec:1}). In light of our review, we discuss possible use cases of the cross-blockchain framework in Section \ref{sec:2}. In Section \ref{open}, we discuss the possible issues that might rise from the employment of the cross-blockchain framework. Finally, we conclude the paper by highlighting challenges and open issues in this research direction. 

\section{Current landscape of Cross blockchain interoperability }\label{Xchain}

Fueled by the unprecedented success of blockchain technology in enabling a decentralized cryptocurrency market, many industries have shown great interest (real and hype) in this technology. In the UAV networks context, researchers, as well as manufacturers, have called for a reimplementation of current UAV networks-based functions to exploit the security advantages of the blockchain and smart contracts. Nonetheless, the omnipresence of these blockchain-based functions has led to fragmentation, redundancies, and fraudulent activities.  Particularly, the implementation of such functions usually takes place on private and thus isolated blockchains. 
Therefore, the interoperability of these isolated blockchains and their associated functions is crucial for enabling a fully connected Internet of blockchains that reduce the overall friction between participating businesses and achieves portability, scalability, and privacy. Moreover, cross-blockchain interoperability is envisioned to reduce redundant transactions and the associated cost. This is especially fundamental for UAV networks which are known to be limited in terms of energy, processing power, and memory \cite{lipton2021blockchain}. A multitude of frameworks has been proposed to achieve blockchain interoperability. Table \ref{IoB} provides a summary of the most common ones.

\begin{table*}[]
\caption{Blockchain interoperability approaches}
\label{IoB}
\begin{tabularx}\textwidth{XXXX}
\hline
Interoperability approach & Description & Weaknesses & Examples \\ \hline
Sidechains & A blockchain connected to the mainchain via CCC protocol. & \begin{tabular}[c]{@{}l@{}}-Require   TTP\\ -Complex   to build \\ and not scalable\end{tabular} & Solana \cite{yakovenko2018solana}, polkadot \cite{wood2016polkadot}, loom \cite{Loom}, RSK \cite{lerner2015rsk}, Horizon \cite{horizonblockchaingames} \\ \hline
Notary schemes & Centralized or decentralized exchanges that make changes to the blockchains on   behalf of users. They are faster and easier   to use. & \begin{tabular}[c]{@{}l@{}}-Mainly   for asset \\ transfers\\ -Most   are centralized\end{tabular} & Binance \cite{binance}, coinbase \cite{coinbase},   kraken \cite{Kraken}\\ \hline
Hashed time-lock contracts & Asset transfer mechanism without a TTP. It locks the fund on one blockchain for a specific time. Fund is unlocked again using a shared secret between the sender and receiver. & \begin{tabular}[c]{@{}l@{}} \\ \\-Sender and receiver\\  need to be online.\\ -Only supports asset\\  transfer.\\ -A secret needs to be\\  created for each   use\end{tabular} & Uniswap \cite{uniswap}, lightning network \cite{lightning}, onechain \cite{onechain}, fusion \cite{fusion} \\ \hline
Blockchain of blockchains & Use case specific blockchains   that interact with each other & \begin{tabular}[c]{@{}l@{}}-Interoperability between \\ blockchains of the\\ same architectures \\ -other architectures \\ are not supported\end{tabular} & -Ethereum   2.0 \cite{ethereum},  Cardano \cite{cardano}, Polkadot \cite{wood2016polkadot}, cosmos \cite{kwon2019cosmos}\\ \hline
Trusted relays & Relay transactions from source   blockchain to destination blockchain & -Private blockchains only & Hyberledger   cactus  \cite{hyperledger} \\ \hline
Blockchain agnostic protocol & Translator between blockchains & \begin{tabular}[c]{@{}l@{}}-Private and public \\ -Some lack enforcing \\ smart contracts and   \\ non-fungible tokens \\ -Rely on TTP\end{tabular} & Quant \cite{quant}, hyberledger Quilt \cite{hyperledger}, interledger \cite{interledger}\\ \hline

\end{tabularx}
\end{table*}
Originally proposed by Back et al. \cite{back2014enabling}, sidechains are secondary blockchains connected to other blockchains via a two-way peg protocol \cite{singh2020sidechain}. This protocol requires locking the transferred funds on the mainchain until they are created in the sidechain. Then, these locked funds may be destroyed. Both, the sidechain and the mainchain may not have the same features or consensus mechanism. However, creating and maintaining sidechains is a complex task because sidechains are designed to interconnect two chains only. Connecting N blockchains requires creating N-1 sidechains, which limits the scalability of this solution. 

Notary schemes are usually centralized exchanges that transfer assets between multiple blockchains. Sometimes, a group of exchanges performs the asset transfer task which is referred to as decentralized notary \cite{schulte2019towards}. Although this approach is the easiest and most convenient, it is prone to centralization-related security risks as well as single-point of failures. 

Hashed time-lock contracts are used for atomic swaps and off-chain transactions between trustless parties. The tokens are locked for a specific time on one blockchain. The receiver can unlock the tokens using by revealing a secret which is shared with him by the sender \cite{decker2015fast}. It, thus, requires sharing a secret between the sender and receiver which may be associated with security risks. Also, it requires the sender and receiver to be online during the transfer time. This is somehow similar to the one-time password (OTP). For this reason, it cannot be considered as a robust interoperability solution in the long run.

Blockchain of blockchains is a framework that connects multiple blockchains in a way similar to sidechains called bridges. Each blockchain is connected to other blockchains in the network either directly or via hubs. The current implementation requires interconnected blockchains to have the same architecture. In addition, this interoperability framework requires additional transaction fees which may prevent scalability on the global level \cite{belchior2020survey}. 

Trusted relay is a decentralized approach that allows validators from source and target chains to validate, sign and deliver transactions between two blockchains. Sometimes, a TTP is employed to perform the tasks of the decentralized verifiers. For instance, Cactus implements multiple TTPs to issue transactions in several blockchains \cite{hegnauer2019design}.

Ideologically, blockchain agnostic protocol is an abstraction layer that allows one to build an application that is operable on multiple blockchains in a seamless manner. Unlike other solutions which depend on constructing bridges between different blockchains, agnostic protocols must be able to function on a higher level layer abstracting from chain-specific protocols. Yet, this solution did not mature to achieve fully interoperable blockchains and is still under development. Although several approaches are proposed \cite{pupyshev2020gravity}, a consensus on a fully agnostic protocol is yet to be unveiled.

A stimulating analogy between the internet and the blockchain has been discussed in \cite{hardjono2019toward}. The authors highlighted the importance of understanding the aspects of the internet that have made it scalable, resilient, sustainable, and commercially viable. They emphasized blockchain interoperability as a crucial requirement for managing and maintaining current and future blockchains. It is argued that the interoperability of the internet is what made it scalable to the global level. Compared to the Internet, blockchains are viewed as Autonomous Systems (AS) that have predefined physical perimeters and are operated by an ISP. 

\section{State-of-the-art on blockchain-based UAV networks}
\label{sec:1}

\begin{figure*}
\centering
  \includegraphics[width=0.8\textwidth]{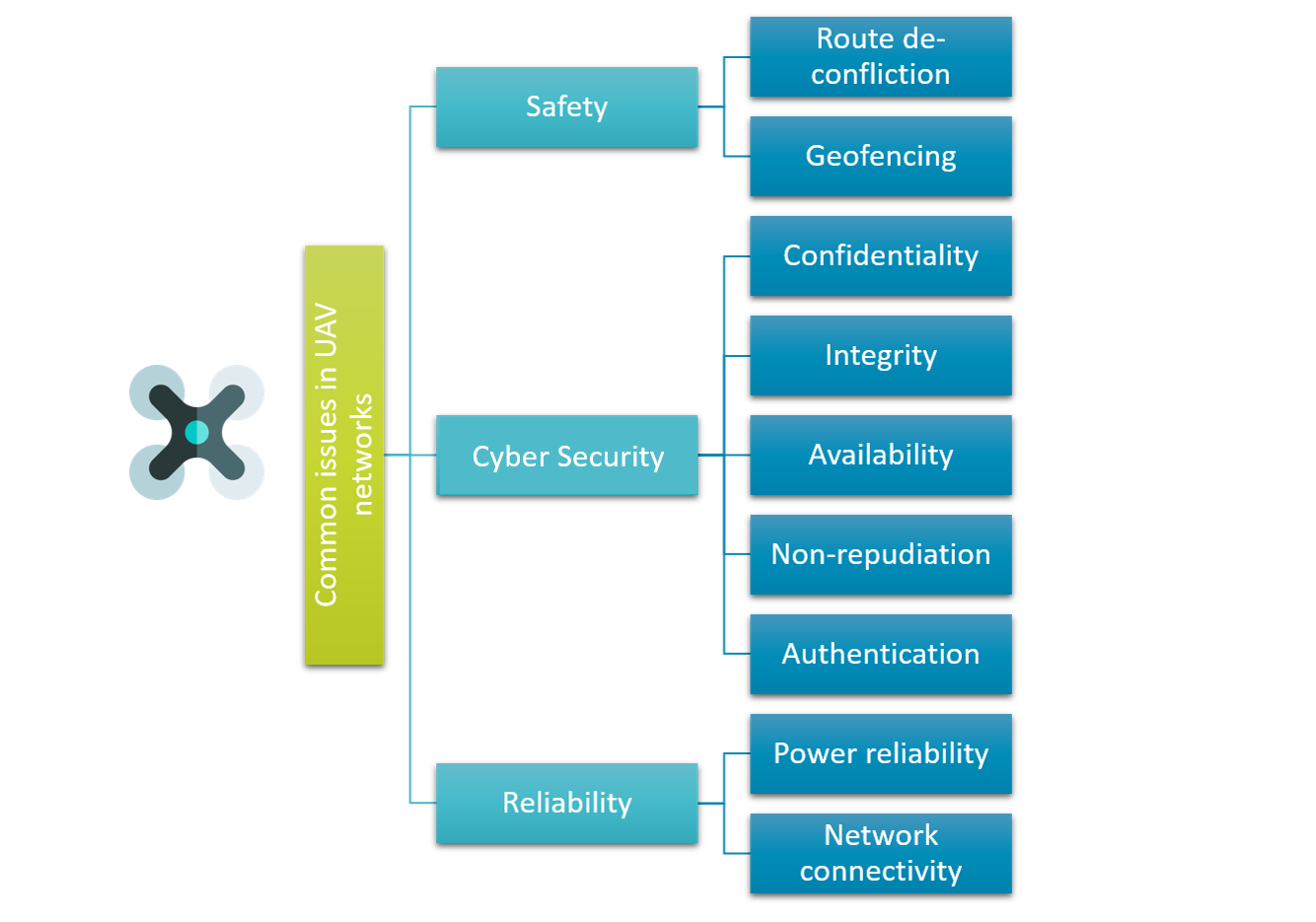}
\caption{Emerging issues in UAV networks that are addressed using blockchain}
\label{fig:1}       
\end{figure*}

The realm of blockchain technology has sufficiently matured to solve emerging  issues related to UAV traffic networks. In this section, we briefly review the state-of-the-art in UAV networks that deploy different versions of blockchain to exploit multiple advantages such as immutability, transparency, traceability, and auditability. Unlike other application-oriented reviews \cite{alladi2020applications}, our review spins three axes based on the ultimate motivation of each work, namely: safety, security, and reliability. We dedicate one subsection for each category to highlight the recent major contributions (from 2018 to 2021). Our review hierarchy is illustrated in Figure \ref{fig:1}. Besides, the focus of the reviewed papers is summarized in Table \ref{tab:1}.

\begin{table*}


\centering
\label{tab:1}       
\begin{tabular}{lll}
\hline\noalign{\smallskip}
Authors & Field of contribution & Blockchain Application   \\
\noalign{\smallskip}\hline\noalign{\smallskip}
Scarlato \textit{et al}. \cite{scarlato2019blockchain} & Safety & Collision avoidance   \\
Rahman \textit{et al}. \cite{rahman} & Safety & Collision avoidance   \\
Kuzmin \textit{et al}. \cite{kuzmin2018blockchain} & Safety & Route deconfliction   \\
Allouch \textit{et al}. \cite{allouch2021utm} & Safety & Route deconfliction   \\

Dasu \textit{et al}. \cite{dasu2018geofences} & Safety & geofencing  \\
Wu \textit{et al}. \cite{wu2020blockchain} & Cybersecurity & Confidentiality: ID management system  \\
 Qian \textit{et al}. \cite{qian2020blockchain} & Cybersecurity & Confidentiality: privacy of cached content  \\
  Xiao \textit{et al}. \cite{xiao2021blockchain} & Cybersecurity & Confidentiality: privacy of cached content  \\
   Ghribi \textit{et al}. \cite{ghribi2020secure} & Cybersecurity & Confidentiality: privacy of cached content  \\
    Lv \textit{et al}. \cite{lv2021analysis} & Cybersecurity & Confidentiality: privacy of cached content  \\
 Islam \textit{et al}. \cite{islam2019buav} & Cybersecurity & Integrity  \\
 Singh \textit{et al}. \cite{singh2020deep} & Cybersecurity & Integrity  \\
 Kanade \textit{et al}. \cite{kanade2021securing} & Cybersecurity & Integrity  \\
  Yazdinejad \textit{et al}. \cite{yazdinejad2020enabling} & Cybersecurity & Availability  \\
  Kapitonov \textit{et al}. \cite{kapitonov2019robonomics} & Cybersecurity & Availability  \\
    Barka \textit{et al}. \cite{barka2019towards} & Cybersecurity & Non-repudiation  \\ 
    Almotary \textit{et al}. \cite{ossamah2020blockchain} & Cybersecurity & Non-repudiation  \\
  Kumari \textit{et al}. \cite{kumari2020taxonomy} & Cybersecurity & Authentication  \\
   Cheema \textit{et al}. \cite{9187206} & Cybersecurity & Authentication   \\
Andola \textit{et al}. \cite{andola2021spychain} & Cybersecurity & Authentication   \\  
  Hassija \textit{et al} \cite{hassija2020distributed} & Reliability & Power reliability: recharging/ refueling  \\
  Jiang \textit{et al} \cite{jiang2020incentivizing} & Reliability & Power reliability  \\
  Qiu \textit{et al} \cite{qiu2019blockchain} & Reliability & Power reliability  \\
  
  Pathak \textit{et al} \cite{pathak2021aerialblocks} & Reliability & Power reliability  \\
  Gai \textit{et al} \cite{gai2020blockchain} & Reliability & Power reliability  \\
   Wu \textit{et al}. \cite{wu2020convergence} & Reliability & Power reliability  \\

 Hassija \textit{et al}  \cite{hassija2020blockchain} & Reliability & Network connectivity  \\
Feng \textit{et al.} \cite{feng2021efficient} & Reliability & Network connectivity  \\
 
  Sharma \textit{et al}. \cite{sharma2019neural} & Reliability & Network coverage  \\

  Aloqaily \textit{et al}. \cite{aloqaily2020design} & Reliability & Network connectivity  \\ 
  
    Singh \textit{et al}. \cite{singh2020odob} & Reliability & Network connectivity  \\ 
  
\noalign{\smallskip}\hline
\end{tabular}
\caption{Latest proposals on blockchain-enabled UAV networks}
\end{table*}

\subsection{Safety}  
 Safety refers to maintaining a good physical condition of a cyber-physical system while in operation. Not only the physical condition of the participating drones shall be preserved, but also the safety of the public residing under the national airspace. This is one of the ultimate goals of a UAV traffic management system that are achieved using different techniques such as by route de-confliction, collision Avoidance, and geofencing.
 
 In principle, \textit{route de-confliction} refers to the mutual planning of flight paths of UAVs in space and time to ensure minimum or no conflicts. Scarlato \textit{et al}. \cite{scarlato2019blockchain} proposed the design of a permissioned blockchain for the collision avoidance and recovery of UAVs. They envisioned a cooperative environment where participating UAVs communicate obstacle coordinates and collisions continuously. \textcolor{black}{Moreover, Rahman \textit{et al}. \cite{rahman} proposed a UAV network to ensure a collision-free environment. Routes are planned in a way to avoid restricted areas such as private properties. Also, the flight altitude is specified to reduce the collision risk by minimizing the number of drones flying at the same height. To ensure that the drone is following the coordinates of the specified route, the authors use a smart contract to log drone movement and location information
during the entire mission. If any of those attributes violates the specified flight route, a negative point is added to the drone's reputation.}
 On the other hand, Kuzmin \textit{et al}. \cite{kuzmin2018blockchain} proposed a route-sharing scheme where cooperative drones deconflict their routes autonomously using the route information on a blockchain. They emphasized the usefulness of this approach especially when a manually operated drone loses connection with its base station. \textcolor{black}{Motivated by a similar application, Allouch \textit{et al.} \cite{allouch2021utm} proposed, implemented, and evaluated a permissioned blockchain to perform secure path planning and data sharing among participating drones. To deal with the limited computation power and storage resources of the UAVs, they offload the computations to a cloud server while employing a decentralized off-chain storage system, namely OrbitDB. Moreover, they exclude the participating UAVs from the peer-to-peer network and only consider ground control stations as peers that store a copy of the ledger. To evaluate their architecture, they implemented the solution on the Hyper-ledger Fabric platform. Finally, they estimated the delay and resource consumption of one transaction. The average latency of an invoke transaction on a network of 50 users was 454 ms. 
Despite being relatively high compared to existing networks delays, this work has shown promise of the application of blockchain-based UAV networks in the real-time. 
}

In contrast, \textit{dynamic geofencing} refers to the virtual geographical fences usually imposed, maintained, and updated by an airspace authority. Dasu \textit{et al}. \cite{dasu2018geofences} presented a hybrid method where parts of the airspace traffic is controlled by a central authority while others are decentralized with the help of blockchain principles. They separate the two parts using dynamic geofencing. In the decentralized zones, participating drones reserve a volume of air to conduct their missions. The reservation is logged in a transaction on a public ledger and is approved if the requested volume is idle at the time of the mission. To achive this, they employ the double-spending avoidance concept originally deployed in cryptocurrencies. The authors also suggest that, depending on the congestion of the required airspace, authorities may charge users for airspace allocation. This approach guarantees a fair share of the airspace while reducing the congestion in peak-hours and urban areas. It also improves the public safety by avoiding collisions and dynamically prohibiting missions over people crowds.   
\subsection{Cybersecurity}
\label{sec:2}
Cybersecurity serves as an overarching goal for the employment of blockchain technology in the UAV networks context. Essentially, the immutability feature of this technology makes it a perfect candidate for ensuring accountability of end-users. Most of the papers that proposed a blockchain-based solution for UAV traffic are motivated by cybersecurity-related requirements, including confidentiality, integrity, availability, non-repudiation, and authentication. We briefly review how each aspect is addressed in the literature. Although, most of the proposed solutions simultaneously address the five security aspects, we tend to group them based on the most dominant one. For a more detailed review of earlier works, the reader is referred to \cite{mehta2020blockchain}.

\textit{Confidentiality} refers to protecting information from being accessed by unauthorized users. Like other networks, UAV networks are prone to confidentiality attacks such as data sniffing, eavesdropping, and replay attacks. Wu \textit{et al}. \cite{wu2020blockchain} have recently outlined multiple scenarios at which blockchain can be employed to preserve the privacy of UAV networks. They suggested a cost-effective, tamper-proof blockchain-based ID management system to authenticate and authorize drones as per the Federal Aviation Authority (FAA) requirements. Besides, they leveraged the potentials of the DLT to preserve the privacy of the trajectory information of the drones. The authors argued that asymmetric encryption and homomorphic obfuscations schemes inherited in blockchain can be utilized to improve the confidentiality of the network. On the other hand, Qian \textit{et al}. \cite{qian2020blockchain} employed blockchain to protect the privacy of cached content by sharing only necessary content with selected vehicles. \textcolor{black}{Similarly, Xiao \textit{et al.} \cite{xiao2021blockchain} proposed a drone-swarm-aided distributed crowd monitoring system that features efficient identity authentication, secure communication, and distributed data management. The goal of the work is to ensure that the monitoring data are kept confidential and secure. They outline a publich key infrastructure (PKI)-based security protocol to authenticate participating UAVs, assign the monitoring task to participating UAVs, and allow secure access for the monitoring data by the management. 
They propose the use of a private blockchain with smart contracts to log the events and actions throughout the task. In terms of results, their simulation revealed that the time overhead of the security protocol increases as the size of the swarm increase, whereas the key creation stage is responsible for the largest delay among other operations.
}

\textcolor{black}{In the same context, Ghribi \textit{et al.} \cite{ghribi2020secure} targeted the confidentiality aspect of UAV networks by implementing a private blockchain with encrypted transactions. The popular public key cryptography with the Elliptic Curve Diffie-Hellman (ECDH) model was used along with a one-time password (OTP). A communication between the sender and the receiver UAVs is done by a nominated endorsement UAV which generates a 128-bit key by using the ECDH. After that, an OTP key is generated by hashing the ECDH key. Finally, the generated key is sent to the sender UAV and the rest of the endorsement UAVs for approval. On the contrary, Lv \textit{et al.} \cite{lv2021analysis} addressed the confidentiality of blockchain aided UAV-networks data sharing scheme by means of the number theory research unit cryptosystem. They argue that their scheme requires low computing cost for encryption, decryption and key generation compared to current data sharing models. 
}

\textcolor{black}{Data \textit{integrity} is also important to ensure that transmitted data is not altered or modified by an intruder.  Islam \textit{et al}. \cite{islam2019buav} proposed an architecture to preserve the integrity of the data transmitted between IoT devices and Mobile Edge Computing (MEC) devices and servers. They used UAVs as trusted relays to ensure the integrity of the data before being transmitted to MEC servers. The data is securely kept in a blockchain at MEC servers. The proposed mechanism which includes an encryption scheme improves the overall integrity of the transferred data and reduces the number of direct requests to MEC servers. In contrast, Singh \textit{et al}. \cite{singh2020deep} employed a blockchain to improve the integrity of data transferred between drones in an IoD environment. Their mechanism intelligently selects the miner node using a deep Boltzmann machine. } \textcolor{black}{
Power plants surveillance is another application that requires high data integrity. Kanade \textit{et al.} \cite{kanade2021securing} proposed a secure surveillance system for power plants using UAV swarms. They employ the DLT to preserve the integrity of the plant data sensed by the UAVs. To reduce the computational cost, transactions are only issued for high risk signals. Further, block validation task is assigned to human workers to reduce the computational cost. 
}

Moreover, maintaining the \textit{availability} of services for UAVs in the airspace is a vital concern. By design, blockchain is a decentralized technology which is immune to single-point-of-failure \cite{agrawal2018continuous}. All proposed blockchain-based UAV traffic management (UTM) architectures are essentially decentralized and require limited or no central authority. Yazdinejad \textit{et al}. \cite{yazdinejad2020enabling} proposed a decentralized zone-based system for registering and authenticating drones. In their architecture, they assign a trusted ground-based drone-controller agent to manage the authentication within a predefined perimeter. They maintain the availability of the authentication scheme by allowing neighbor drone-controllers to substitute for a failing one. \textcolor{black}{In the UTM context, the authors of \cite{kapitonov2019robonomics} defined a blockchain architecture for UTM in which they implemented a so called Robonomic protocol to provide security between communicating nodes. They argue that their architecture solves the latency issue that is found in decentralized systems, which is essentially beneficial for critical applications. The model is supported by smart contracts that provide transparency and immutability. Thus, the work is implemented by means of the decentralized Ethereum and InterPlanetary File System.}

Another crucial requirement for the cybersecurity of UAV networks is \textit{non-repudiation}. This term is defined as the inability to deny or refuse responsibilities of actions. This can be achieved using public key infrastructure, where a UAV signs messages using its private key before sending it via the network. In \cite{barka2019towards}, Barka \textit{et al}. highlighted the importance of non-repudiation as a requirement for UAV networks. They argued that a UAV might deny sending images of restricted areas. Thus, the authors proposed a trusted blockchain-based UAV system to protect critical infrastructure. \textcolor{black}{ Similarly, the authors of \cite{ossamah2020blockchain} defined four blockchain concepts that can enhance the drone security. These elements comprise: digital fingerprint, data structure, conensus mechanism, and access control. They emphasize the role of the consensus mechanism in preventing dishonesty by allowing nodes in the network to agree on a transaction.} 

Finally, ensuring \textit{authenticity} of users and messages is a key requirement for UAV networks. In principle, authentication is defined as the ability to recognize the real identity of a user. UAV networks may suffer from authentication attacks such as masquerading attack. Kumari \textit{et al}. \cite{kumari2020taxonomy} addressed such threats in UAV networks by presenting a blockchain-based softwarization of the UAV network management scheme. They argue that the UAV network should be provisioned with cryptographic data to ensure authentication and confidentiality. Similarly, Cheema \textit{et al}. \cite{9187206} employed blockchain technology to develop a registration and authentication scheme for drones in a smart vehicular network context. They also addressed the issue of optimal drone positioning to enhance the overall spectral efficiency in the network. \textcolor{black}{Under the same consideration, Andola \textit{et al.} \cite{andola2021spychain} proposed a lightweight blockchain model that provides authentication and anonymity. Their work provides security preferences for a surveillance UAV. They defined an adversary model in which an attacker can modify the transactions of a blockchain during the communication before they are verified. This kind of attack is called  a malleability attack. They also defined an issue during the handoff process when a UAV moves from one Ground Control station (GCS) to another, where latency might occur. They solve this issue by implementing a novel blockchain architecture which they called SpyChain. They have described a detailed authentication scheme of their work following four approaches}. 

\subsection{Reliability}
Improving the reliability of UAV traffic management is a crucial prerequisite to attracting large-scale commercial applications such as package delivery, transportation, and network coverage. This is achieved by optimizing power reliability and network connectivity while minimizing the cost. A surge of research activities have been identified in this direction. 

Hassija \textit{et al} \cite{hassija2020distributed} presented a blockchain-based architecture to manage power-charging/refueling between UAVs and charging stations. Their proposed model allows UAVs to buy power using digital tokens. Wireless power trading in UAV networks based on blockchain principles is also proposed in \cite{jiang2020incentivizing}. \textcolor{black}{The same idea was also proposed earlier in \cite{qiu2019blockchain} where a consortium blockchain was exploited to introduce a spectrum trading platform for UAV-assisted cellular networks. In this work, MEC devices were employed to reduce the computation overhead of the verification process. To select the block miner, they implemented a reputation algorithm for the MEC devices. Further, the business model is incentivised  by a Stackelberg game to maximize the profit of the cellular network operators and the UAV operators. They claim that the proposed framework enables a secure, efficient, and decentralized spectrum trading between both trading parties. Another UAV-based business model is proposed by Pathak \textit{et al.} \cite{pathak2021aerialblocks} where a blockchain-based UAV virtualization is introduced to provide UAV as a service. The envisioned platform aims at connecting UAV owners, end users, and UAV-service providers via a permissioned distributed network, enabling UAV owners to rent their vehicles to end users which in turns allow for a more efficient utilization of resources. Their architecture is similar to the previously proposed sensor virtualization \cite{misra2019qos}.  The concept of blockchain is employed to ensure efficient resource utilization, security, and market competition.  Comparably, Gai \textit{et al.} \cite{gai2020blockchain} incentivised miners to validate authentication and authorization certificates. The presented approach aims at facilitating secure and reliable group communication between UAVs. Particularly, blockchain was employed to record and validate actions. On the other hand, an attribute-based voting mechanism is introduced by means of smart contracts. Real-time experiments were carried out to verify and assess the model performance. On the other hand, Wu \textit{et al}. \cite{wu2020convergence} proposed a layered IoT architecture that supports offloading computationally-intensive mining processes to edge servers. This, in turn, reduces power consumption on UAVs and base stations. } 

To improve the 5G network coverage, the authors of \cite{hassija2020blockchain} proposed another blockchain-based model for using drones as dynamic base-stations. Their model integrates a game-theoretic smart contract that ensures fair and efficient allocation of bandwidth between users. \textcolor{black}{Likewise, the authors of \cite{feng2021efficient} proposed a blockchain enabled 5G drones network to address the identity authentication and secure data sharing of drones. In their work, they implement three core services utilizing blockcahin technology, namely: identity authentication, operation management, and security auditing. Moreover, they employ a multi-signature smart contract managed by a central authority for registering and authenticating drones. 
To incentivise miners, peers request a certain amount of coins to verify that the requesting drone is already in the registered drones list. Besides, secure data sharing is enabled by uploading encrypted data to the cloud. Similarly, the uploading process is mainly managed by a smart contract deployed on the blockchain. 
}
On the contrary, Sharma \textit{et al}. \cite{sharma2019neural} presented a neural-blockchain-based scheme for MEC caching. In their model, they used drones as base stations to provide ultra-reliable flattened 5G network service. The proposed model was evaluated in terms of flyby time and area spectral efficiency. Similarly, Aloqaily \textit{et al}. \cite{aloqaily2020design} envisioned a blockchain assisted 5G network which improves the quality of service by deploying public and private ledgers supported by fog and cloud data centers. They showed that their framework improves packet delivery success rate compared to traditional networks without blockchain. \textcolor{black}{Additionally, Singh \textit{et al.} \cite{singh2020odob} addressed the problem of reliability in UAV-networks by introducing a light-weight permissioned blockchain solution that in which each drone in the network would access its own block rather than all the blocks in the ledger. They argue that this architecture solves the issue of large ledger that would be produced each time more data and blocks are added. A shrinking mechanism has also been implemented to provide a fast lightweight blockchain. }
\section{Cross-blockchain-enabled Framework for UAV Networks}
\begin{figure*}
\centering
  \includegraphics[width=0.8\textwidth]{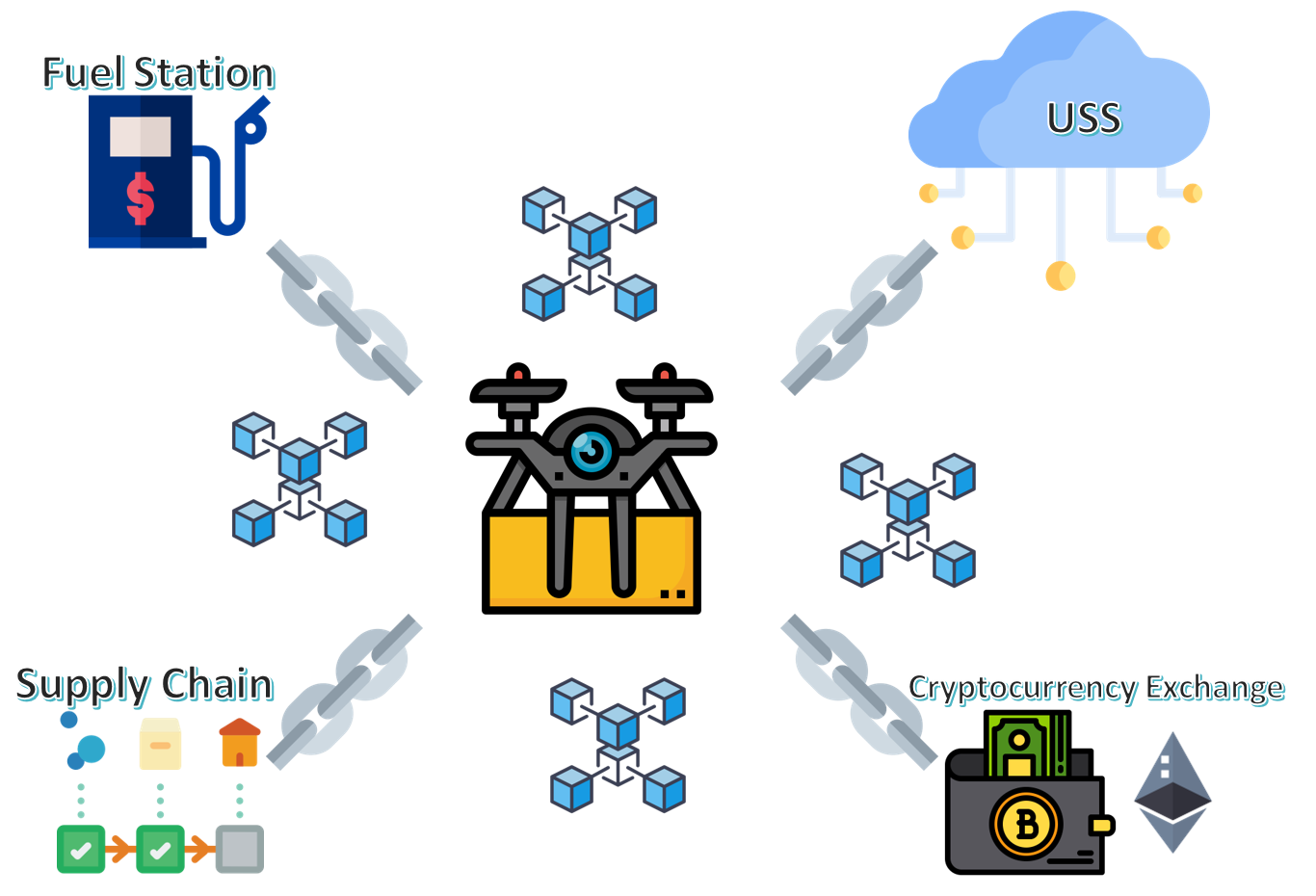}
\caption{Potential cross-blockchain scenario at which a delivery drone is operating on four ledgers simultaneously.}
\label{fig:2}       
\end{figure*}
Despite the remarkable advances in blockchain-aided UAV networks, the exploitation of cross-blockchain communication framework is still limited. The main purpose of the cross-blockchain technology is to connect the \textit{independent} blockchain networks. A variety of solutions have been introduced to cope with the interoperability limitation. \textcolor{black}{Qasse \textit{et al.} categorized cross blockchain solutions into four groups: sidechains \cite{poon2017plasma}, trusted third party \cite{wood2016polkadot}, blockchain routers \cite{ding2018interchain,kan2018multiple}, and smart contracts \cite{bennink2018analysis}. Alternatively, Belchoir \textit{et al.} \cite{belchior2020survey} categorized the proposed approaches into three categories, comprising: public connectors \cite{wang2020electricity,rueegger2020rational}, hybrid connectors \cite{abebe2019enabling,fynn2020smart}, and blockchain of blockchains \cite{spoke2017aion,kwon2019cosmos}.   As discussed in the previous section, blockchain is playing an important role in building UAV networks. Basically, UAV networks have a variety of functionalities, yet each paper discussed focuses on either a single or double functionalities. Eventually, a sophisticated network must combine all of the functionalities to build a secure and safe UAV environment.} \textcolor{black}{One could take lessons from previously implemented cross-blockchain models in other fields such as asset transfer \cite{sigwart2020decentralized,borkowski2018caught} and health records \cite{cao2020ceps}. Fueled by the enormous applications that may potentially benefit from the cross-blockchain concept, software development environment were also introduced to enable easier deployment of cross-blockchain models \cite{qiu2019chainide}, as well as the associated smart contracts \cite{nissl2020towards}.  }In this section, we highlight opportunities and discuss multiple proposals to employ interoperable blockchains in the IoD environment including: multiple UAV Service Suppliers (USS), UAV charging stations, and UAV delivery application (supply chain). 

\subsection{Multiple UAV Service Suppliers}
The USS is an entity that provides services to subscribed UAS operators to help them meet the operational requirements specified by the national aviation authority. Operation planning, strategic and tactical de-confliction, Remote ID (RID), and airspace authorization are examples of the services provided by a USS. A UAV operator may subscribe to one or more USSs to avail multiple services. Upon subscription to a USS, the UAV is given a unique ID and automatically registered on a public blockchain which is accessible by other USSs and aviation authorities. In parallel, the USS deliver digital assets/data to the UAV using another blockchain that is only visible to the UAVs in a certain zone and this particular USS. The USS needs to interface the two ledgers to synchronize the process of serving new subscribers while keeping record of their IDs in a public blockchain.

Another scenario could happen when a USS user migrates to another USS. In this case, he might request the old USS to migrate his data and reputation/awards to the new USS to make use of them. Recall that keeping all assets (i.e. registration, payment, reputation, IDs, location-based services, etc.) in one ledger is inefficient, especially when the chain becomes too long. Moreover, some of these information shall not be shared with some parties while other should be public. Thus, adopting specialized ledgers and allowing communication between them improves scalability and confidentiality. 

On the other hand, UAV networks could also make use of the decentralized identifiers (DID) scheme to enable the minimum disclosure of users' information on a need-to-know basis. Many DID frameworks are built on top of a blockchain \cite{w3c}. Indeed, DID preserves users' privacy while promoting a universally unique identity that can be used across multiple blockchains \cite{kim2021security}. Currently, the DID framework is being considered for universal standardization by the World Wide Web Consortium (W3C) \cite{alzahrani2020information} which implies the importance of integrating it with the current blockchain-based UAV networks. Chen et al \cite{chen2021bidm} proposed a decentralized cross-domain blockchain that interfaces multiple applications via a DID system. Ideologically, UAV networks could make use of such solutions after carefully tailoring them for the UAV environment. 

\subsection{UAV charging stations}
As discussed earlier, Hassija \textit{et al}. \cite{hassija2020distributed} envisioned a UAV charging-refueling scheme built on top of the blockchain. They assumed that UAVs can buy power/fuel from ground stations using tokens or cryptocurrencies. However, every fuel supplier might restrict transactions on its blockchain using a particular cryptocurrencies. What if the UAV does not hold coins of the same cryptocurrency in its wallet? A straightforward solution would be to implement cross-blockchain solutions to exchange coins to the desired cryptocurrency. 
\subsection{Physical Assets Delivery}
The need to incorporate cross-blockchain technology in UAV networks becomes evident in the physical-assets delivery scenario. That is, a delivery UAV needs to operate on at least two blockchains: one related to the supply-chain, and another related to the airspace traffic network. In some cases, the supply-chain ledger might request traffic and location-related information from the air traffic ledger to track the shipment. Possibly, this use case could be addressed using the CAPER framework \cite{amiri2019caper}, where confidentiality and interoperability are jointly provided. In other cases, the two ledgers may need to exchange coins to pay for USS services and refueling. Figure \ref{fig:2} illustrates a potential scenario at which blockchain interoperability is required for a delivery drone. In this scenario, four blockchains need to interoperate to accomplish a simple delivery mission. In the first blockchain, the drone subscribes to a USS smart contract at which it gets flight-related services such as dynamic geofencing and route de-confliction. To subscribe to this USS, a specific cryptocurrency wallet is required, which lives in another ledger. Further, the UAV needs to connect to a private supply chain related ledger, where the supplier and costumer can track the shipment location. Another possible use of blockchain is for fueling or charging the drone using a re-charging station. Essentially, the drone needs to exchange information securely between these ledgers without compromising data integrity or security. A cross-blockchain solution is indispensable in such cases.

\section{Open issues and challenges} \label{open}

Both, blockchain interoperability and UAV networks are still in the research and development phase. Thus, the integration of both technologies is faced with many practical problems. We devote this section to discuss challenges and possible research opportunities that may guide the future work in this topic.  
\paragraph{Selection of cross-blockchain technology: }
Despite the substantial efforts dedicated to develop efficient and secure cross-blockchain interfaces \cite{belchior2020survey}, the selection of the most appropriate one for managing the different applications in the UAV networks is a critical design challenge. All currently available cross-blockchain platforms still suffer from either security, reliability, or efficiency issues. Consequently, it is inevitable to choose the optimum platform that minimizes the risks while maintaining satisfactory performance. 
\paragraph{Absence of a managing third party}
Although the decentralized nature of peer-to-peer (P2P) networks provides enormous advantages and mitigates substantial security risks, it is, yet, vulnerable to 51\% attacks \cite{bb}. This is especially intimidating in the case of UAV networks, where UAVs can join and leave freely \cite{wu2020blockchain}. Also, the wide availability of relatively cheap UAVs may facilitate such attacks. To tackle this issue, Dasu \textit{et al}. \cite{dasu2018geofences} presented a hybrid traffic management model where parts of the airspace are managed by a centralized authority, while others are left to blockchain-based decentralized management. Further, the authors presented a novel approach to prevent denial-of-airspace attack by charging airspace users some fees based on the local demand. This will also limit 51\% attacks as it makes it more expensive for malicious users to join the UAV network with many drones at a time. 
\textcolor{black}{Another way to improve the security feature of the cross-blockchain scheme was introduced by Kim \textit{et al.} \cite{kim2019trailer,kim2021enhanced,kim2019new} where the blockchain governance game was proposed to deal with attackers who try to gain control over the blockchain by adding more illegitimate nodes. }
\paragraph{Cost effectiveness:}
Proposing any cross-blockchain solution in the UAV network management context shall be cost effective. That is, most of the civilian UAV-based applications are commercial and thus the successful deployment of any solution in the drone market is significantly dependent on the associated costs. 

\paragraph{Simulation tools:}

We emphasize the importance of developing effective simulation tools for blockchain-based UAV applications, where the computational costs, memory size, message overhead, and UAV dynamics are taken into consideration. The availability of such tools will definitely drive the development of efficient cross-blockchain solutions tailored for the unique dynamic nature of the IoD environment. Moreover, Mehta \textit{et al}. \cite{mehta2020blockchain} highlighted the impact of the lack of proper programming practices on the efficiency of the blockchain implementations. 

\paragraph{Limited computational and storage resources:}

This is the major barrier in the application of blockchain in the UAV networks. This fact is emphasized in \cite{wu2020blockchain,alladi2020applications,mehta2020blockchain}. Particularly, most current consensus mechanisms are, by design, power-consuming. Also, keeping a copy of the blockchain on the UAV board requires a large memory.  In fact, increasing the capacity and computational power of UAVs while minimizing power consumption is one of the most active research topics. Yet, a game-changing solution is far from being deployed. Thus, it will be helpful to implement lightweight consensus mechanism such as the one presented in \cite{9035635}. Relying on edge serves for computational-intensive tasks and storage is also proposed \cite{wu2020convergence}. Nonetheless, one should pay attention to the associated communication overheads and trust issues. 

\paragraph{Compatibility of Blockchain Ledgers:}
By reviewing the related literature, it had come to our notice that each research uses a different type of blockchain. The work done by \cite{rahman} focused on the permissioned blockchain to ensure the authorization of blockchain miners. Yet,  security vulnerabilities may arise if other blockchain ledgers in the cross-blockchain platform employ the permissionless architecture. This might lead to reducing the security level that was granted by the permissioned blockchain. In other words, certain blockchains might be implemented either as public or private, permissioned or permissionless. If combined together, the security level of the entire system will be determined by the least-secure component. Another issue that requires further research is whether or not combining different blockchain ledgers will introduce new vulnerabilities to a system. 

\section{Conclusion}
The blockchain technology sees to offer promising features in many applications. In this paper we discussed different employments of the blockchain to advance the safety, cybersecurity, and reliability of UAV networks. We identified different properties that has been addressed and exploited in recent literature and concluded that the deployment of this technology is faced with many barriers such as scalability and portability. A range of cross-blockchain interoperability solutions has been proposed to improve scalability while maintaining transparency, immutability, and decentralization. Motivated by the advances in cross-blockchain solutions, we outlined multiple scenarios at which current blockchain-based UAV networks may potentially profit from the deployment of blockchain inter-operation protocols. This includes a scenario in which the functionalities of four different blockchain ledgers are combined to fulfill the purpose of a single UAV system operating under the concept of cross-blockchain.  We then, highlighted the challenges associated with implementing a network of blockchains to concurrently enable multiple functionalities of UAV networks. Some of these challenges include cost, computational and storage resources, and the compatibility of different blockchain ledgers. We finally suggested possible research directions that will form the basis for future proposals to integrate cross-blockchain solutions in the IoD environment.


%

%



\begin{thebibliography}{10}

\bibitem{binance}
Binance.
\newblock https://www.binance.com/en.
\newblock Accessed: 2021-07-11.

\bibitem{cardano}
Cardano.
\newblock http://www.cardano.org/.
\newblock Accessed: 2021-07-11.

\bibitem{coinbase}
Coinbase.
\newblock https://www.coinbase.com/.
\newblock Accessed: 2021-07-11.

\bibitem{w3c}
Decentralized identifiers (dids) v1.0.

\bibitem{ethereum}
Ethereum 2.0.
\newblock https://ethereum.org/en/.
\newblock Accessed: 2021-07-11.

\bibitem{horizonblockchaingames}
Horizon blockchain games.
\newblock https://horizon.io/.
\newblock Accessed: 2021-07-11.

\bibitem{interledger}
Interledger.
\newblock
  https://interledger.org/news/attention-creatives-grant-for-the-webs-call-for-proposals-is-now-open/.

\bibitem{Loom}
Intro to loom network: Loom sdk.
\newblock {https://loomx.io/developers/en/intro-to-loom.html}.

\bibitem{Kraken}
Kraken exchange.
\newblock https://www.kraken.com/.
\newblock Accessed: 2021-07-11.

\bibitem{lightning}
Lightning network.
\newblock https://lightning.network/.
\newblock Accessed: 2021-07-11.

\bibitem{onechain}
Onechain.
\newblock $http://www.onechain.one/index_en.html.$
\newblock Accessed: 2021-07-11.

\bibitem{quant}
Quant.
\newblock https://www.quant.network/.
\newblock Accessed: 2021-07-11.

\bibitem{uniswap}
Uniswap.
\newblock https://uniswap.org/.
\newblock Accessed: 2021-07-11.

\bibitem{hyperledger}
Hyperledger: open source blockchain technologies.
\newblock https://www.hyperledger.org/, Oct 2021.

\bibitem{abebe2019enabling}
Ermyas Abebe, Dushyant Behl, Chander Govindarajan, Yining Hu, Dileban
  Karunamoorthy, Petr Novotny, Vinayaka Pandit, Venkatraman Ramakrishna, and
  Christian Vecchiola.
\newblock Enabling enterprise blockchain interoperability with trusted data
  transfer (industry track).
\newblock In {\em Proceedings of the 20th International Middleware Conference
  Industrial Track}, pages 29--35, 2019.

\bibitem{agrawal2018continuous}
Rahul Agrawal, Pratik Verma, Rahul Sonanis, Umang Goel, Aloknath De,
  Sai~Anirudh Kondaveeti, and Suman Shekhar.
\newblock Continuous security in iot using blockchain.
\newblock In {\em 2018 IEEE International Conference on Acoustics, Speech and
  Signal Processing (ICASSP)}, pages 6423--6427. IEEE, 2018.

\bibitem{alladi2020applications}
Tejasvi Alladi, Vinay Chamola, Nishad Sahu, and Mohsen Guizani.
\newblock Applications of blockchain in unmanned aerial vehicles: A review.
\newblock {\em Vehicular Communications}, page 100249, 2020.

\bibitem{allouch2021utm}
Azza Allouch, Omar Cheikhrouhou, Anis Koub{\^a}a, Khalifa Toumi, Mohamed
  Khalgui, and Tuan Nguyen~Gia.
\newblock Utm-chain: blockchain-based secure unmanned traffic management for
  internet of drones.
\newblock {\em Sensors}, 21(9):3049, 2021.

\bibitem{aloqaily2020design}
Moayad Aloqaily, Ouns Bouachir, Azzedine Boukerche, and Ismaeel~Al Ridhawi.
\newblock Design guidelines for blockchain-assisted 5g-uav networks.
\newblock {\em arXiv preprint arXiv:2007.15286}, 2020.

\bibitem{alzahrani2020information}
Bander Alzahrani.
\newblock An information-centric networking based registry for decentralized
  identifiers and verifiable credentials.
\newblock {\em IEEE Access}, 8:137198--137208, 2020.

\bibitem{amiri2019caper}
Mohammad~Javad Amiri, Divyakant Agrawal, and Amr~El Abbadi.
\newblock Caper: a cross-application permissioned blockchain.
\newblock {\em Proceedings of the VLDB Endowment}, 12(11):1385--1398, 2019.

\bibitem{andola2021spychain}
Nitish Andola, Vijay~Kumar Yadav, S~Venkatesan, Shekhar Verma, et~al.
\newblock Spychain: A lightweight blockchain for authentication and anonymous
  authorization in iod.
\newblock {\em Wireless Personal Communications}, pages 1--20, 2021.

\bibitem{back2014enabling}
Adam Back, Matt Corallo, Luke Dashjr, Mark Friedenbach, Gregory Maxwell, Andrew
  Miller, Andrew Poelstra, Jorge Tim{\'o}n, and Pieter Wuille.
\newblock Enabling blockchain innovations with pegged sidechains.
\newblock {\em URL: http://www. opensciencereview.
  com/papers/123/enablingblockchain-innovations-with-pegged-sidechains}, 72,
  2014.

\bibitem{barka2019towards}
Ezedin Barka, Chaker~Abdelaziz Kerrache, Hadjer Benkraouda, Khaled Shuaib,
  Farhan Ahmad, and Fatih Kurugollu.
\newblock Towards a trusted unmanned aerial system using blockchain for the
  protection of critical infrastructure.
\newblock {\em Transactions on Emerging Telecommunications Technologies}, page
  e3706, 2019.

\bibitem{belchior2020survey}
Rafael Belchior, Andr{\'e} Vasconcelos, S{\'e}rgio Guerreiro, and Miguel
  Correia.
\newblock A survey on blockchain interoperability: Past, present, and future
  trends.
\newblock {\em arXiv preprint arXiv:2005.14282}, 2020.

\bibitem{bennink2018analysis}
Peter Bennink, Lennart~van Gijtenbeek, Oskar~van Deventer, and Maarten Everts.
\newblock An analysis of atomic swaps on and between ethereum blockchains using
  smart contracts.
\newblock Technical report, Tech. report, 2018.

\bibitem{besanccon2019towards}
L{\'e}o Besan{\c{c}}on, Catarina~Ferreira Da~Silva, and Parisa Ghodous.
\newblock Towards blockchain interoperability: Improving video games data
  exchange.
\newblock In {\em 2019 IEEE International Conference on Blockchain and
  Cryptocurrency (ICBC)}, pages 81--85. IEEE, 2019.

\bibitem{med}
S.~{Biswas}, K.~{Sharif}, F.~{Li}, Z.~{Latif}, S.~S. {Kanhere}, and S.~P.
  {Mohanty}.
\newblock Interoperability and synchronization management of blockchain-based
  decentralized e-health systems.
\newblock {\em IEEE Transactions on Engineering Management}, 67(4):1363--1376,
  2020.

\bibitem{borkowski2018caught}
Michael Borkowski, Christoph Ritzer, Daniel McDonald, and Stefan Schulte.
\newblock Caught in chains: claim-first transactions for cross-blockchain asset
  transfers.
\newblock {\em TU Wien: Technische Universit{\"a}t Wien, Tech. Rep}, 2018.

\bibitem{cao2020ceps}
Sheng Cao, Jing Wang, Xiaojiang Du, Xiaosong Zhang, and Xiaolin Qin.
\newblock Ceps: A cross-blockchain based electronic health records
  privacy-preserving scheme.
\newblock In {\em ICC 2020-2020 IEEE International Conference on Communications
  (ICC)}, pages 1--6. IEEE, 2020.

\bibitem{9187206}
M.~A. {Cheema}, M.~K. {Shehzad}, H.~K. {Qureshi}, S.~A. {Hassan}, and
  H.~{Jung}.
\newblock A drone-aided blockchain-based smart vehicular network.
\newblock {\em IEEE Transactions on Intelligent Transportation Systems}, pages
  1--11, 2020.

\bibitem{chen2021bidm}
Ruibiao Chen, Fangxing Shu, Shuokang Huang, Lei Huang, Huafang Liu, Jin Liu,
  and Kai Lei.
\newblock Bidm: A blockchain-enabled cross-domain identity management system.
\newblock {\em Journal of Communications and Information Networks},
  6(1):44--58, 2021.

\bibitem{dasu2018geofences}
Tamraparni Dasu, Yaron Kanza, and Divesh Srivastava.
\newblock Geofences in the sky: herding drones with blockchains and 5g.
\newblock In {\em Proceedings of the 26th ACM SIGSPATIAL International
  Conference on Advances in Geographic Information Systems}, pages 73--76,
  2018.

\bibitem{decker2015fast}
Christian Decker and Roger Wattenhofer.
\newblock A fast and scalable payment network with bitcoin duplex micropayment
  channels.
\newblock In {\em Symposium on Self-Stabilizing Systems}, pages 3--18.
  Springer, 2015.

\bibitem{ding2018interchain}
Donghui Ding, Tiantian Duan, Linpeng Jia, Kang Li, Zhongcheng Li, and Yi~Sun.
\newblock Interchain: A framework to support blockchain interoperability.
\newblock {\em Second Asia-Pacific Work. Netw}, 2018.

\bibitem{feng2021efficient}
Chaosheng Feng, Keping Yu, Ali~Kashif Bashir, Yasser~D Al-Otaibi, Yang Lu,
  Shengbo Chen, and Di~Zhang.
\newblock Efficient and secure data sharing for 5g flying drones: a
  blockchain-enabled approach.
\newblock {\em IEEE Network}, 35(1):130--137, 2021.

\bibitem{fusion}
Fusion.
\newblock A connected ecosystem for financial transactions.
\newblock https://fusion.org/en.
\newblock Accessed: 2021-07-11.

\bibitem{fynn2020smart}
Enrique Fynn, Alysson Bessani, and Fernando Pedone.
\newblock Smart contracts on the move.
\newblock In {\em 2020 50th Annual IEEE/IFIP International Conference on
  Dependable Systems and Networks (DSN)}, pages 233--244. IEEE, 2020.

\bibitem{gai2020blockchain}
Keke Gai, Yulu Wu, Liehuang Zhu, Kim-Kwang~Raymond Choo, and Bin Xiao.
\newblock Blockchain-enabled trustworthy group communications in uav networks.
\newblock {\em IEEE Transactions on Intelligent Transportation Systems}, 2020.

\bibitem{gharibi2016internet}
Mirmojtaba Gharibi, Raouf Boutaba, and Steven~L Waslander.
\newblock Internet of drones.
\newblock {\em IEEE Access}, 4:1148--1162, 2016.

\bibitem{ghribi2020secure}
Elias Ghribi, Tala~Talaei Khoei, Hamed~Taheri Gorji, Prakash Ranganathan, and
  Naima Kaabouch.
\newblock A secure blockchain-based communication approach for uav networks.
\newblock In {\em 2020 IEEE International Conference on Electro Information
  Technology (EIT)}, pages 411--415. IEEE, 2020.

\bibitem{hardjono2019toward}
Thomas Hardjono, Alexander Lipton, and Alex Pentland.
\newblock Toward an interoperability architecture for blockchain autonomous
  systems.
\newblock {\em IEEE Transactions on Engineering Management}, 67(4):1298--1309,
  2019.

\bibitem{hassija2020distributed}
Vikas Hassija, Vinay Chamola, Dara Nanda~Gopala Krishna, and Mohsen Guizani.
\newblock A distributed framework for energy trading between uavs and charging
  stations for critical applications.
\newblock {\em IEEE Transactions on Vehicular Technology}, 69(5):5391--5402,
  2020.

\bibitem{hassija2020blockchain}
Vikas Hassija, Vikas Saxena, and Vinay Chamola.
\newblock A blockchain-based framework for drone-mounted base stations in
  tactile internet environment.
\newblock In {\em IEEE INFOCOM 2020-IEEE Conference on Computer Communications
  Workshops (INFOCOM WKSHPS)}, pages 261--266. IEEE, 2020.

\bibitem{hegnauer2019design}
Timo Hegnauer.
\newblock {\em Design and development of a blockchain interoperability api}.
\newblock PhD thesis, Master’s thesis, CSG@ IFI, University of Zurich,
  Switzerland, to appear 2019~…, 2019.

\bibitem{furtune}
Furtune~Business Insights.
\newblock Small drones market.
\newblock {\em
  https://www.fortunebusinessinsights.com/press-release/small-drones-market-9611},
  21-02-2020.

\bibitem{islam2019buav}
Anik Islam and Soo~Young Shin.
\newblock Buav: A blockchain based secure uav-assisted data acquisition scheme
  in internet of things.
\newblock {\em Journal of Communications and Networks}, 21(5):491--502, 2019.

\bibitem{jiang2020incentivizing}
Li~Jiang, Bin Chen, Shengli Xie, Sabita Maharjan, and Yan Zhang.
\newblock Incentivizing resource cooperation for blockchain empowered wireless
  power transfer in uav networks.
\newblock {\em IEEE Transactions on Vehicular Technology}, 2020.

\bibitem{kan2018multiple}
Luo Kan, Yu~Wei, Amjad~Hafiz Muhammad, Wang Siyuan, Gao Linchao, and Hu~Kai.
\newblock A multiple blockchains architecture on inter-blockchain
  communication.
\newblock In {\em 2018 IEEE International Conference on Software Quality,
  Reliability and Security Companion (QRS-C)}, pages 139--145. IEEE, 2018.

\bibitem{kanade2021securing}
Vijay~A Kanade.
\newblock Securing drone-based ad hoc network using blockchain.
\newblock In {\em 2021 International Conference on Artificial Intelligence and
  Smart Systems (ICAIS)}, pages 1314--1318. IEEE, 2021.

\bibitem{kapitonov2019robonomics}
Aleksandr Kapitonov, Ivan Berman, Vadim Manaenko, Vyacheslav Rzhevskiy, Vitaly
  Bulatov, and Artemii Zenkin.
\newblock Robonomics as a blockchain-based platform for unmanned traffic
  management of mobile vehicles.
\newblock In {\em 2019 Workshop on Research, Education and Development of
  Unmanned Aerial Systems (RED UAS)}, pages 9--17. IEEE, 2019.

\bibitem{khan2019extended}
Muhammad~Yasar Khan, Megat~F Zuhairi, Toqeer Ali, Turki Alghamdi, and
  Jose~Antonio Marmolejo-Saucedo.
\newblock An extended access control model for permissioned blockchain
  frameworks.
\newblock {\em Wireless Networks}, pages 1--12, 2019.

\bibitem{kim2021security}
Bong~Gon Kim, Young-Seob Cho, Seok-Hyun Kim, Hyoungshick Kim, and Simon~S Woo.
\newblock A security analysis of blockchain-based did services.
\newblock {\em IEEE Access}, 9:22894--22913, 2021.

\bibitem{kim2019trailer}
Song-Kyoo Kim.
\newblock The trailer of strategic alliance for blockchain governance game.
\newblock {\em arXiv preprint arXiv:1903.11172}, 2019.

\bibitem{kim2019new}
Song-Kyoo Kim, Chan~Yeob Yeun, Ernesto Damiani, Yousef Al-Hammadi, and Nai-Wei
  Lo.
\newblock New blockchain adoptation for automotive security by using systematic
  innovation.
\newblock In {\em 2019 IEEE Transportation Electrification Conference and Expo,
  Asia-Pacific (ITEC Asia-Pacific)}, pages 1--4. IEEE, 2019.

\bibitem{kim2021enhanced}
Song-Kyoo~Amang Kim.
\newblock Enhanced iov security network by using blockchain governance game.
\newblock {\em Mathematics}, 9(2):109, 2021.

\bibitem{kumari2020taxonomy}
Aparna Kumari, Rajesh Gupta, Sudeep Tanwar, and Neeraj Kumar.
\newblock A taxonomy of blockchain-enabled softwarization for secure uav
  network.
\newblock {\em Computer Communications}, 161:304--323, 2020.

\bibitem{kuzmin2018blockchain}
Alexander Kuzmin and Evgeny Znak.
\newblock Blockchain-base structures for a secure and operate network of
  semi-autonomous unmanned aerial vehicles.
\newblock In {\em 2018 IEEE International Conference on Service Operations and
  Logistics, and Informatics (SOLI)}, pages 32--37. IEEE, 2018.

\bibitem{kwon2019cosmos}
Jae Kwon and Ethan Buchman.
\newblock Cosmos whitepaper, 2019.

\bibitem{lerner2015rsk}
Sergio~Demian Lerner.
\newblock Rsk.
\newblock 2015.

\bibitem{lipton2021blockchain}
A~Lipton and T~Hardjono.
\newblock Blockchain intra-and interoperability.
\newblock {\em Innovative Technology at the Interface of Finance and
  Operations. Springer}, 2021.

\bibitem{lv2021analysis}
Zhihan Lv, Liang Qiao, M~Shamim Hossain, and Bong~Jun Choi.
\newblock Analysis of using blockchain to protect the privacy of drone big
  data.
\newblock {\em IEEE Network}, 35(1):44--49, 2021.

\bibitem{mehta2020blockchain}
Parimal Mehta, Rajesh Gupta, and Sudeep Tanwar.
\newblock Blockchain envisioned uav networks: Challenges, solutions, and
  comparisons.
\newblock {\em Computer Communications}, 151:518--538, 2020.

\bibitem{misra2019qos}
Sudip Misra and Aishwariya Chakraborty.
\newblock Qos-aware dispersed dynamic mapping of virtual sensors in
  sensor-cloud.
\newblock {\em IEEE Transactions on Services Computing}, 2019.

\bibitem{neisse2020interledger}
Ricardo Neisse, Jos{\'e}~L Hern{\'a}ndez-Ramos, Sara~N Matheu-Garcia, Gianmarco
  Baldini, Antonio Skarmeta, Vasilios Siris, Dmitrij Lagutin, and Pekka
  Nikander.
\newblock An interledger blockchain platform for cross-border management of
  cybersecurity information.
\newblock {\em IEEE Internet Computing}, 24(3):19--29, 2020.

\bibitem{nissl2020towards}
Markus Nissl, Emanuel Sallinger, Stefan Schulte, and Michael Borkowski.
\newblock Towards cross-blockchain smart contracts.
\newblock {\em arXiv preprint arXiv:2010.07352}, 2020.

\bibitem{ossamah2020blockchain}
Almotery Ossamah.
\newblock Blockchain as a solution to drone cybersecurity.
\newblock In {\em 2020 IEEE 6th World Forum on Internet of Things (WF-IoT)},
  pages 1--9. IEEE, 2020.

\bibitem{pathak2021aerialblocks}
Nidhi Pathak, Anandarup Mukherjee, and Sudip Misra.
\newblock Aerialblocks: Blockchain-enabled uav virtualization for industrial
  iot.
\newblock {\em IEEE Internet of Things Magazine}, 4(1):72--77, 2021.

\bibitem{poon2017plasma}
Joseph Poon and Vitalik Buterin.
\newblock Plasma: Scalable autonomous smart contracts.
\newblock {\em White paper}, pages 1--47, 2017.

\bibitem{pupyshev2020gravity}
Aleksei Pupyshev, Dmitry Gubanov, Elshan Dzhafarov, Ilya Sapranidi, Inal
  Kardanov, Vladimir Zhuravlev, Shamil Khalilov, Marc Jansen, Sten Laureyssens,
  Igor Pavlov, et~al.
\newblock Gravity: a blockchain-agnostic cross-chain communication and data
  oracles protocol.
\newblock {\em arXiv preprint arXiv:2007.00966}, 2020.

\bibitem{qian2020blockchain}
Yongfeng Qian, Yingying Jiang, Long Hu, M~Shamim Hossain, Mubarak Alrashoud,
  and Muneer Al-Hammadi.
\newblock Blockchain-based privacy-aware content caching in cognitive internet
  of vehicles.
\newblock {\em IEEE Network}, 34(2):46--51, 2020.

\bibitem{qiu2019chainide}
Han Qiu, Xiao Wu, Shuyi Zhang, Victor~CM Leung, and Wei Cai.
\newblock Chainide: A cloud-based integrated development environment for
  cross-blockchain smart contracts.
\newblock In {\em 2019 IEEE International Conference on Cloud Computing
  Technology and Science (CloudCom)}, pages 317--319. IEEE Computer Society,
  2019.

\bibitem{qiu2019blockchain}
Junfei Qiu, David Grace, Guoru Ding, Junnan Yao, and Qihui Wu.
\newblock Blockchain-based secure spectrum trading for
  unmanned-aerial-vehicle-assisted cellular networks: An operator’s
  perspective.
\newblock {\em IEEE Internet of Things Journal}, 7(1):451--466, 2019.

\bibitem{rahman}
Mohammad~Saidur Rahman, Ibrahim Khalil, and Mohammed Atiquzzaman.
\newblock Blockchain-powered policy enforcement for ensuring flight compliance
  in drone-based service systems.
\newblock {\em IEEE Network}, 35(1):116--123, 2021.

\bibitem{rueegger2020rational}
Janick Rueegger and Guilherme~Sperb Machado.
\newblock Rational exchange: Incentives in atomic cross chain swaps.
\newblock In {\em 2020 IEEE International Conference on Blockchain and
  Cryptocurrency (ICBC)}, pages 1--3. IEEE, 2020.

\bibitem{bb}
Muhammad Saad, Jeffrey Spaulding, Laurent Njilla, Charles~A Kamhoua, D~Nyang,
  and Aziz Mohaisen.
\newblock Overview of attack surfaces in blockchain.
\newblock {\em Blockchain for distributed systems security}, pages 51--66,
  2019.

\bibitem{scarlato2019blockchain}
Michele Scarlato, Cristian Perra, Mohamed~Yaseen Jabarulla, Giljun Jung, and
  Heung~No Lee.
\newblock A blockchain for the collision avoidance and the recovery of crashed
  uavs.
\newblock pages 463--467, 2019.

\bibitem{schulte2019towards}
Stefan Schulte, Marten Sigwart, Philipp Frauenthaler, and Michael Borkowski.
\newblock Towards blockchain interoperability.
\newblock In {\em International conference on business process management},
  pages 3--10. Springer, 2019.

\bibitem{sharma2019neural}
Vishal Sharma, Ilsun You, Dushantha Nalin~K Jayakody, Daniel~Gutierrez Reina,
  and Kim-Kwang~Raymond Choo.
\newblock Neural-blockchain-based ultrareliable caching for edge-enabled uav
  networks.
\newblock {\em IEEE Transactions on Industrial Informatics}, 15(10):5723--5736,
  2019.

\bibitem{sigwart2020decentralized}
Marten Sigwart, Philipp Frauenthaler, Christof Spanring, and Stefan Schulte.
\newblock Decentralized cross-blockchain asset transfers.
\newblock {\em arXiv preprint arXiv:2004.10488}, 2020.

\bibitem{singh2020sidechain}
Amritraj Singh, Kelly Click, Reza~M Parizi, Qi~Zhang, Ali Dehghantanha, and
  Kim-Kwang~Raymond Choo.
\newblock Sidechain technologies in blockchain networks: An examination and
  state-of-the-art review.
\newblock {\em Journal of Network and Computer Applications}, 149:102471, 2020.

\bibitem{singh2020deep}
Maninderpal Singh, Gagangeet~Singh Aujla, and Rasmeet~Singh Bali.
\newblock A deep learning-based blockchain mechanism for secure internet of
  drones environment.
\newblock {\em IEEE Transactions on Intelligent Transportation Systems}, 2020.

\bibitem{singh2020odob}
Maninderpal Singh, Gagangeet~Singh Aujla, and Rasmeet~Singh Bali.
\newblock Odob: One drone one block-based lightweight blockchain architecture
  for internet of drones.
\newblock In {\em IEEE INFOCOM 2020-IEEE Conference on Computer Communications
  Workshops (INFOCOM WKSHPS)}, pages 249--254. IEEE, 2020.

\bibitem{spoke2017aion}
Matthew Spoke, NE~Team, et~al.
\newblock Aion: Enabling the decentralized internet.
\newblock {\em AION, White Paper, Jul}, 2017.

\bibitem{9035635}
Z.~{Su}, Y.~{Wang}, Q.~{Xu}, and N.~{Zhang}.
\newblock Lvbs: Lightweight vehicular blockchain for secure data sharing in
  disaster rescue.
\newblock {\em IEEE Transactions on Dependable and Secure Computing}, pages
  1--1, 2020.

\bibitem{wang2020electricity}
Hongkai Wang, Dong He, Xiaoyi Wang, Caichao Xu, Weiwei Qiu, Yiyang Yao, and
  Qiang Wang.
\newblock An electricity cross-chain platform based on sidechain relay.
\newblock In {\em Journal of Physics: Conference Series}, volume 1631, page
  012189. IOP Publishing, 2020.

\bibitem{wood2016polkadot}
Gavin Wood.
\newblock Polkadot: Vision for a heterogeneous multi-chain framework.
\newblock {\em White Paper}, 21, 2016.

\bibitem{wu2020convergence}
Yulei Wu, Hong-Ning Dai, and Hao Wang.
\newblock Convergence of blockchain and edge computing for secure and scalable
  iiot critical infrastructures in industry 4.0.
\newblock {\em IEEE Internet of Things Journal}, 2020.

\bibitem{wu2020blockchain}
Yulei Wu, Hong-Ning Dai, Hao Wang, and Kim-Kwang~Raymond Choo.
\newblock Blockchain-based privacy preservation for 5g-enabled drone
  communications.
\newblock {\em arXiv preprint arXiv:2009.03164}, 2020.

\bibitem{xiao2021blockchain}
Wenjing Xiao, Miao Li, Bander Alzahrani, Reem Alotaibi, Ahmed Barnawi, and
  Qingsong Ai.
\newblock A blockchain-based secure crowd monitoring system using uav swarm.
\newblock {\em IEEE Network}, 35(1):108--115, 2021.

\bibitem{yakovenko2018solana}
Anatoly Yakovenko.
\newblock Solana: A new architecture for a high performance blockchain v0.
  8.13.
\newblock {\em Whitepaper}, 2018.

\bibitem{yazdinejad2020enabling}
Abbas Yazdinejad, Reza~M Parizi, Ali Dehghantanha, Hadis Karimipour, Gautam
  Srivastava, and Mohammed Aledhari.
\newblock Enabling drones in the internet of things with decentralized
  blockchain-based security.
\newblock {\em IEEE Internet of Things Journal}, 2020.

\end{thebibliography}


%
%

\end{document}